# Projective measurement in nuclear magnetic resonance

Jae-Seung Lee and A. K. Khitrin

Department of Chemistry, Kent State University, Kent, Ohio 44242-0001

**Abstract**

It is demonstrated that nuclear magnetic resonance experiments using pseudopure spin states can give possible outcomes of projective quantum measurement and probabilities of such outcomes. The physical system is a cluster of six dipolar-coupled nuclear spins of benzene in a liquid-crystalline matrix. For this system with the maximum total spin $S=3$, the results of measuring $S_X$ are presented for the cases when the state of the system is one of the eigenstates of $S_Z$.

Interesting quantum algorithms, like factorization of numbers [1] or teleportation [2], rely on projective measurement and collapse of wave function. Today, clusters of nuclear spins are the largest and the most complex quantum systems where individual states have been addressed. Using nuclear magnetic resonance (NMR) technique, manipulations with individual quantum states have been demonstrated for systems of up to twelve qubits [3-5]. Conventional detection of signal in NMR is a so-called ensemble weak measurement, when weak interaction with radio-frequency coil does not change significantly quantum states of individual spins or molecules in the process of measuring the total magnetization of a sample. Possibility to retrieve from NMR experiments the results of projective measurements will enhance capabilities of NMR as an inexpensive tool for testing quantum algorithms. In this Letter we show that NMR experiments with pseudopure spin states can be designed to provide information about possible outcomes of projective measurement and probabilities of such outcomes.

The experiments have been performed on clusters of six dipolar-coupled proton spins of benzene in liquid crystal ZLI-1167 at 25°C using Varian Unity/Inova 500 MHz NMR spectrometer. This system is a good example of an ensemble of non-interacting spin clusters, where each benzene molecule contains six proton spins, coupled by residual dipole-dipole interactions. The spin Hamiltonian is

$$H = -\omega_0 \sum_{k=1}^{6} S_{kZ} + \sum_{k>j}^{6} b_{jk}\{S_{kZ} S_{jZ} - (1/2) S_{kX} S_{jX} - (1/2) S_{kY} S_{jY}\}, \quad (1)$$

where $\omega_0$ is the Larmor frequency, $S_{kX}$, $S_{kY}$, and $S_{kZ}$ are the components of $k$-th spin, and $b_{jk}$ are the constants of residual dipole-dipole interaction. J-couplings between the proton spins are small, with the largest one $J_{12}/2\pi = 7.5$ Hz [6], and can be neglected. The technique for preparing pseudopure states [7,8] for this system was described elsewhere [9]. The eigenfunctions of the system's Hamiltonian (1) are simultaneously eigenfunctions of $S_Z$, the projection of the total spin on the direction of external magnetic field. We start with one of the eigenfunctions of the Hamiltonian (1) and measure $S_X$, the x-component of the total spin of the cluster in the rotating frame. Since $S_X$ does not commute with $S_Z$, the measurement cannot give a definite value. The amplitudes of probabilities of possible outcomes are projections of the initial state on the eigenstates of $S_X$. To project the state of the system on the $S_X$ eigenstates, we instantly turn on a strong field along x-axis of the rotating frame, as it is shown on the experimental scheme in Fig. 1. This strong field not only locks the total magnetization [10], but it also preserves populations of individual eigenstates of $S_X$ corresponding to different values of the magnetic quantum number $M_X$ (transitions changing $M_X$ are forbidden by the energy conservation). Then, the frequency of the radio-frequency pulse gradually changes so that the direction of the effective field adiabatically rotates towards z-axis. When the frequency offset becomes much greater than the dipole-dipole interactions in the cluster, the amplitude of the radio-frequency pulse is adiabatically decreased to zero. As a result, each of the $S_X$ eigenstates is adiabatically converted into one of the eigenstates of the system's Hamiltonian with the same value of the magnetic quantum number [11]. Populations of these states, obtained from linear-response NMR spectra, give probabilities of different values of $M_X$ for the initial state.

The Hamiltonian (1) does not commute with $S^2$. However, its eigenstates with $M_Z=\pm 3$ and $M_Z=\pm 2$ are also the eigenstates of $S^2$. Two states with all spins up ($M_Z=+3$) or down ($M_Z=-3$) are $S=3$ states. Twelve states with $M_Z=\pm 2$ are spin waves [12]. Two states with the wavenumber $k=0$ are $S=3$ states. Ten other spin waves with $k\neq 0$ are $S=2$ states.

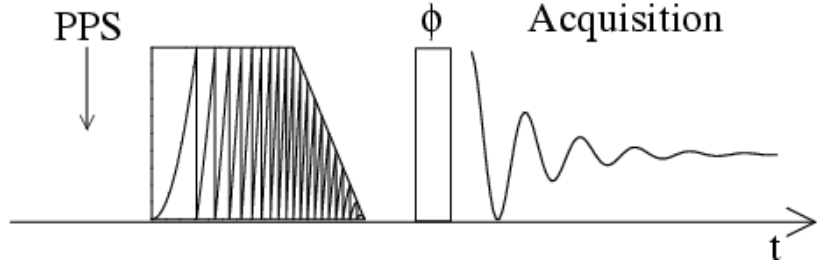

Fig. 1 Experimental pulse sequence: preparation of pseudopure state is followed by a frequency-sweeping lock pulse of 25 ms duration, initial amplitude $\gamma B_1/2\pi$ = 19 kHz, and 20 kHz sweeping during the first 20 ms; a linear-response spectrum is acquired using a small-angle reading pulse.

We start our first experiment with the ($S$=3, $M_Z$=3) pseudopure state. Linear-response spectrum of this state, shown in Fig. 2b, has only one peak corresponding to a single-quantum transition to the ($S$=3, $M_Z$=2) state. Since the initial state is an eigenstate of $S^2$ with $S$=3, it projects on only seven of the $S_X$ eigenstates, which belong to the $S$=3 subspace and can be described by a single quantum number $M_X$ ranging from $-3$ to $+3$. Adiabatic pulse converts each of these seven states into one of the seven high-spin eigenstates of the Hamiltonian (1) with the same magnetic quantum number. The high-spin eigenstates with $M_Z$=±3 and $M_Z$=±2 are $S$=3 states. The states $M_Z$=±1 and $M_Z$=0 are, respectively, the states with effective spins $S$=2.78 and $S$=2.59, calculated as $S(S+1)=<S^2>$. At the same time, each of these high-spin states is the state with the lowest energy among the states with the same value of $M_Z$ [13]. Six single-quantum transitions between seven high-spin eigenstates are marked with asterisks in the thermal-equilibrium spectrum in Fig. 2a. At thermal equilibrium, all differences of populations for pairs of states with $\Delta M_Z$=1 are equal and, therefore, the intensities of the peaks in Fig. 2a are proportional to the transition probabilities (squares of the transition matrix elements).

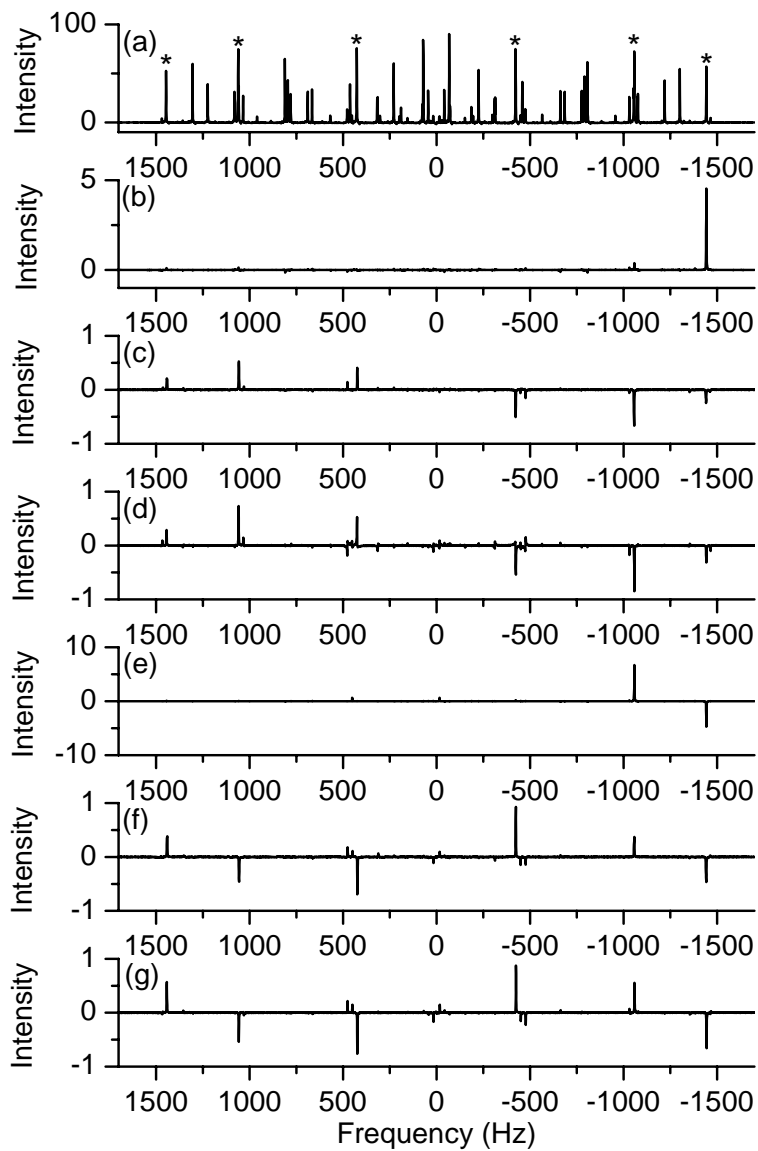

Fig. 2 (a) Thermal equilibrium NMR spectrum, the peaks marked by asterisks are the transitions between seven high-spin states; (b) linear-response spectrum of the ($S$=3, $M_Z$=3) state; (c) linear-response spectrum for the ($S$=3, $M_Z$=3) state projected onto $S_X$ by a lock pulse; (d) linear-response spectrum for the ($S$=3, $M_Z$=3) state after 90° hard pulse and gradient dephasing; (e) linear-response spectrum of the ($S$=3, $M_Z$=2) state; (f) linear-response spectrum for the ($S$=3, $M_Z$=2) state projected onto $S_X$ by a lock pulse; (g) linear-response spectrum for the ($S$=3, $M_Z$=2) state after 90° hard pulse and gradient dephasing.

The result of experiment for the initial ($S=3$, $M_Z=3$) state is shown in Fig. 2c. Analysis of this linear-response spectrum confirms that only seven high-spin eigenstates of the Hamiltonian (1) are populated. Six differences of populations for these seven states are obtained from intensities of six major peaks in Fig. 2c, normalized by the transition probabilities (intensities of the peaks in the thermal equilibrium spectrum in Fig. 2a). The integration constant is obtained from the condition that the sum of populations equals to the population of the pseudopure state in Fig. 2b. Experimental populations of the seven states, which represent probabilities of seven different outcomes of the $M_X$ measurement are shown in Fig. 3a. The theoretical probabilities: 1/64, 6/64, 15/64, 20/64, 15/64, 6/64, 1/64 for $M_X$ ranging from –3 to +3 are also shown in Fig. 3a for comparison. The theoretical value of $\langle S_X^2 \rangle$ for the ($S=3$, $M_Z=3$) state is $\langle S_X^2 \rangle = \langle S_Y^2 \rangle = (1/2)\{S(S+1)-M_Z^2\} = 3/2$. The value calculated with the experimental probability distribution in Fig. 3a is $\langle S_X^2 \rangle = 1.57$.

Another way to obtain the results of projective measurement is to rotate eigenfunctions of the measured operator to the measurement basis. This method has been used in experiments with trapped ions [14] and photons [15]. Linear-response spectrum after applying a 90° hard pulse to the ($S=3$, $M_Z=3$) state and dephasing with magnetic field gradient is shown in Fig. 2d. Gradient dephasing has been used before to imitate the effect of projective measurement in NMR [16]. Although the spectra 2e and 2d look similar, they are not the same. Since there are no $S=3$ eigenstates of the system's Hamiltonian with $M_Z=\pm 1$ and $M_Z=0$, many $M_Z=\pm 1$ and $M_Z=0$ states become populated in the experiment using 90° pulse, which results in additional small peaks in Fig. 2d. In order to accurately determine the experimental probabilities of possible outcomes, one needs to add all populations of the states with corresponding value of $M_Z$. Therefore, for complex systems, adiabatic evolution is more convenient because it does not increase the number of populated states.

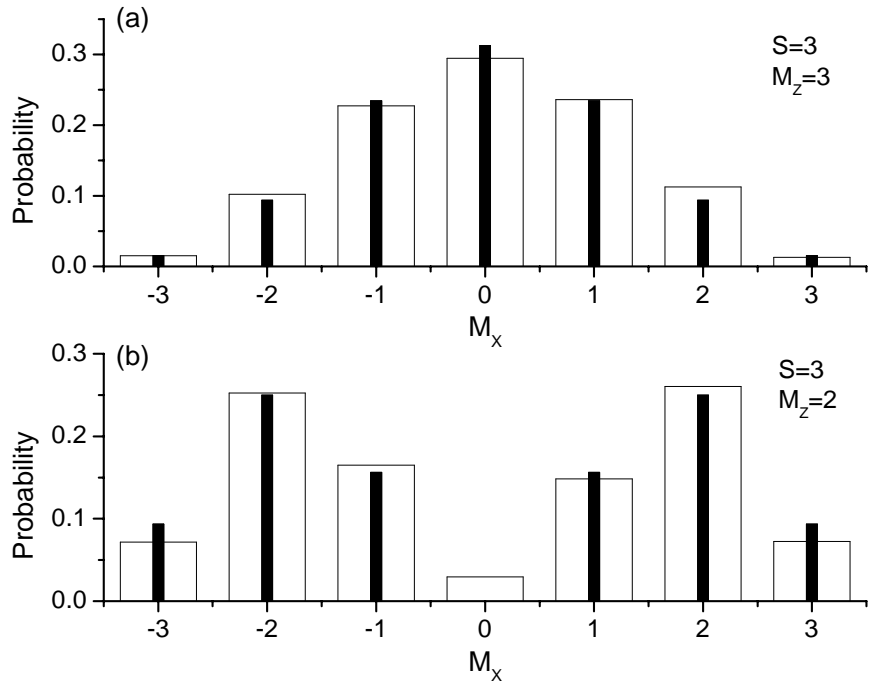

Fig. 3 Probabilities of different values of $M_X$ for the initial (a) ($S=3$, $M_Z=3$) and (b) ($S=3$, $M_Z=2$) states are shown with wide bars; narrow bars are the theoretical values.

Pseudopure eigenstate of the Hamiltonian (1) with $S=3$ and $M_Z=2$ was obtained from the ($S=3$, $M_Z=3$) state by applying a 180° Gaussian selective pulse of 30 ms duration and maximum amplitude $\gamma B_1/2\pi = 15$ Hz at the frequency of the peak in Fig. 2b. The linear-response spectrum for this state is shown in Fig. 2e. It consists of the negative peak, which is the transition to the ($S=3$, $M_Z=3$) state, large positive peak, which is the transition to the high-spin $M_Z=1$ state, and two small positive peaks, barely seen in Fig. 3e, which are the transitions to two other $M_Z=1$ states [17]. For the initial ($S=3$, $M_Z=2$) state, experimental populations of the seven states, which represent probabilities of seven different outcomes of the $M_X$ measurement are shown in Fig. 3b. The theoretical probabilities: 3/32, 8/32, 5/32, 0, 5/32, 8/32, 3/32 for $M_X$ ranging from $-3$ to $+3$ are also shown in Fig. 3b. The theoretical value of $\langle S_X^2 \rangle$ for the state with $S=3$ $M_Z=2$ is $\langle S_X^2 \rangle = \langle S_Y^2 \rangle = (1/2)\{S(S+1)-M_Z^2\} = 4$. The value calculated with the experimental probability distribution in Fig. 3b is $\langle S_X^2 \rangle = 3.66$.

In conclusion, NMR experiments, performed with pseudopure spin states, can provide information about possible outcomes of projective measurement and their probabilities. The technique is based on turning on instantly the Hamiltonian representing the measured quantity and then, adiabatically changing it to another Hamiltonian with spectroscopically distinguishable eigenstates. Different members of an ensemble, representing different outcomes of projective measurement, can be not only identified by their spectra but also individually addressed with frequency-selective pulses. This allows implementing schemes where operations are conditioned by the results of projective measurement.

The work was supported by Kent State University and Ohio Board of Regents.

**References**

[1] P. Shor, Proc. of 35[th] Annual Symposium on the Foundations of Computer Science, IEEE Computer Society, Los Alamitos, CA, p. 124 (1994).

[2] C. H. Bennett, G. Brassard, C. Crépeau, R. Jozsa, A. Peres, and W. K. Wootters, Phys. Rev. Lett. 70, 1895 (1993).

[3] J.-S. Lee and A. K. Khitrin, J. Chem. Phys. 122, 041101 (2005).

[4] J.-S. Lee and A. K. Khitrin, Appl. Phys. Lett. 87, 204109 (2005).

[5] C. Negrevergne, T.S. Mahesh, C. A. Ryan, M. Ditty, F. Cyr-Racine, W. Power, N. Boulant, T. Havel, D.G. Cory, and R. Laflamme, arXiv:quant-ph/0603248

[6] A. K. Khitrin and B. M. Fung, J. Chem. Phys. 111, 7480 (1999).

[7] D. G. Cory, A. F. Fahmy, and T. F. Havel, Proc. Natl. Acad. Sci. U.S.A. 94, 1634 (1997).

[8] N. Gershenfeld and I. L. Chuang, Science 275, 350 (1997).

[9] J.-S. Lee and A. K. Khitrin, Phys. Rev. A 70, 022330 (2004).

[10] M. H. Levitt, *Spin dynamics*, (John Wiley & Sons, Chichester, 2001).

[11] A. Messiah, *Quantum Mechanics* (Wiley, New York, 1976).

[12] E. B. Feldman and A. K. Khitrin, Zh. Eksp. Teor. Fiz. (JETP) 98, 967 (1990).

[13] J.-S. Lee, K. E. Cardwell, and A. K. Khitrin, Phys. Rev. A 72, 064101 (2005).

[14] P. C. Haljan, P. J. Lee, K-A. Brickman, M. Acton, L. Deslauriers, and C. Monroe, Phys. Rev. A 72, 062316 (2005).

[15] K. J. Resch, P. Walther, and A. Zeilinger, Phys. Rev. Lett. 94, 070402 (2005).

[16] J. A. Jones, arXiv:quant-ph/0506235.

[17] A, Saupe, Z. Naturforsch. A 20, 572 (1965).